\RequirePackage[2020-02-02]{latexrelease}
\documentclass[twocolumn, switch]{article} 

\usepackage{preprint}

\usepackage{amsmath, amsthm, amssymb, amsfonts}

\usepackage[numbers,square]{natbib}
\usepackage[utf8]{inputenc}	
\usepackage[T1]{fontenc}	
\usepackage{xcolor}		
\usepackage[colorlinks = true,
            linkcolor = purple,
            urlcolor  = blue,
            citecolor = cyan,
            anchorcolor = black]{hyperref}	
\usepackage{booktabs} 		
\usepackage{nicefrac}		
\usepackage{microtype}		
\usepackage{lineno}		
\usepackage{float}			

\usepackage{graphicx}
\usepackage{textcomp}
\usepackage[justification=centering]{caption}
\usepackage[justification=centering]{subcaption}
\usepackage[ruled,linesnumbered]{algorithm2e}
\usepackage[noend]{algpseudocode}
\usepackage{multirow}
\usepackage{diagbox}
\usepackage{tabularx}
\usepackage{booktabs}
\usepackage{makecell}
\usepackage{url}

\makeatletter
\newcommand{\algorithmfootnote}[2][\footnotesize]{%
  \let\old@algocf@finish\@algocf@finish
  \def\@algocf@finish{\old@algocf@finish
    \leavevmode\rlap{\begin{minipage}{\linewidth}
    #1#2
    \end{minipage}}%
  }%
}
\makeatother

\usepackage{newfloat}
\DeclareFloatingEnvironment[name={Supplementary Figure}]{suppfigure}
\usepackage{sidecap}
\sidecaptionvpos{figure}{c}

\usepackage{titlesec}
\titlespacing\section{0pt}{12pt plus 3pt minus 3pt}{1pt plus 1pt minus 1pt}
\titlespacing\subsection{0pt}{10pt plus 3pt minus 3pt}{1pt plus 1pt minus 1pt}
\titlespacing\subsubsection{0pt}{8pt plus 3pt minus 3pt}{1pt plus 1pt minus 1pt}

\usepackage{tikz,xcolor,hyperref}

\definecolor{lime}{HTML}{A6CE39}
\DeclareRobustCommand{\orcidicon}{
	\begin{tikzpicture}
	\draw[lime, fill=lime] (0,0) 
	circle [radius=0.16] 
	node[white] {{\fontfamily{qag}\selectfont \tiny ID}};
	\draw[white, fill=white] (-0.0625,0.095) 
	circle [radius=0.007];
	\end{tikzpicture}
	\hspace{-2mm}
}
\foreach \x in {A, ..., Z}{\expandafter\xdef\csname orcid\x\endcsname{\noexpand\href{https://orcid.org/\csname orcidauthor\x\endcsname}
			{\noexpand\orcidicon}}
}

\title{A Two-Stage Federated Transfer Learning Framework in Medical Images Classification on Limited Data: A COVID-19 Case Study}

\usepackage{xwatermark}
\newwatermark[firstpage,color=gray!60,angle=90,scale=0.32, xpos=-4.05in,ypos=0]{\href{https://doi.org/}{\color{gray}{Publication doi}}}
\newwatermark[firstpage,color=gray!60,angle=90,scale=0.32, xpos=3.9in,ypos=0]{\href{https://doi.org/}{\color{gray}{Preprint doi}}}
\newwatermark[firstpage,color=gray!90,angle=0,scale=0.28, xpos=0in,ypos=-5in]{*correspondence: \texttt{skzhang@nmsu.edu}}

\usepackage{authblk}

\author[1\thanks{\tt{skzhang@nmsu.edu}}]{Alexandros Shikun Zhang}
\author[2]{Naomi Fengqi Li}

\affil[1]{Department of Computer Science, New Mexico State University, Las Cruces, NM, USA}
\affil[2]{Independent Scholar, Las Cruces, NM, USA}

\begin{document}

\twocolumn[ 
  \begin{@twocolumnfalse} 
  
\maketitle

\begin{abstract}
COVID-19 pandemic has spread rapidly and caused a shortage of global medical resources. The efficiency of COVID-19 diagnosis has become highly significant. As deep learning and convolutional neural network (CNN) has been widely utilized and been verified in analyzing medical images, it has become a powerful tool for computer-assisted diagnosis. However, there are two most significant challenges in medical image classification with the help of deep learning and neural networks, one of them is the difficulty of acquiring enough samples, which may lead to model overfitting. Privacy concerns mainly bring the other challenge since medical-related records are often deemed patients' private information and protected by laws such as GDPR and HIPPA. Federated learning can ensure the model training is decentralized on different devices and no data is shared among them, which guarantees privacy. However, with data located on different devices, the accessible data of each device could be limited. Since transfer learning has been verified in dealing with limited data with good performance, therefore, in this paper, We made a trial to implement federated learning and transfer learning techniques using CNNs to classify COVID-19 using lung CT scans. We also explored the impact of dataset distribution at the client-side in federated learning and the number of training epochs a model is trained. Finally, we obtained very high performance with federated learning, demonstrating our success in leveraging accuracy and privacy.
\end{abstract}


\vspace{0.35cm}

  \end{@twocolumnfalse} 
] 



\section{Introduction}
\label{sec:introduction}
Caused by severe acute respiratory syndrome coronavirus-2, coronavirus disease 2019 (COVID-19) has become an ongoing pandemic after it was found at the end of the year 2019, due to the fast spread and infection rate of the COVID-19 epidemic, the World Health Organization (WHO) designated it a pandemic \citep{world2020laboratory}. It will be critical to find tools, processes, and resources to rapidly identify individuals infected.

According to several previous researches \citep{carotti2020chest, herpe2020efficacy, parry2020chest, ufuk2020chest, tenda2020importance, francone2020chest}, computed tomography (CT) offers a high diagnostic and prognostic value for COVID-19, CT scans of individuals with COVID-19 often revealed bilateral lung lesions comprised of ground-glass opacity \citep{ai2020correlation} and in some cases, abnormalities and changes were observed \citep{lee2020covid}. Since CT scans are a popular diagnostic technique that is simple and quick to get without incurring the significant expense, incorporating CT imaging into the development of a sensitive diagnostic tool may expedite the diagnosis process while also serving as a complement to RT-PCR \citep{xie2020chest, ai2020correlation, gietema2020ct}. However, utilizing CT imaging to forecast a patient's individualized prognosis may identify prospective high-risk individuals who are more likely to develop seriously and need immediate medical attention. Researchers have realized that developing effective methods to assist diagnosis has become critical to their success.

As a key machine learning method, deep learning has evolved in recent years and has achieved astonishing success in the field of medical image processing \citep{shen2017deep, litjens2017survey, haskins2020deep}. Because of the superior capability of convolutional neural networks (CNNs) in medical image classification, researchers have begun to concentrate their attention on the application of CNNs in order to address an increasing number of medical image processing issues using deep learning, and some previous researches have demonstrated the great capability of CNNs being implemented in computed assisted diagnosis \citep{duran2020prometeo, tanaka2019computer, okamoto2019feature, zhang2019computer, barbu2018analysis, qiu2017new, sun2016computer}. Some previous researches have also achieved exciting results in COVID-19 classification \citep{ibrahim2021deep, ibrahim2021pneumonia, hussain2021corodet, sakib2020dl, chen2021design, gupta2021instacovnet, elzeki2021covid, pham2020comprehensive, keidar2021covid, narin2021automatic, gilanie2021coronavirus}, however, since medical records of patients have always been deemed as privacy and protected by laws such as GDPR in the European Union (EU) and HIPPA in the US, collecting data needed for building high-quality classifiers such as CT scans becomes extremely difficult. In some other researches \citep{feki2021federated, yan2021experiments, dayan2021federated, kumar2021blockchain, nguyen2021federated, cetinkaya2021communication, zhang2021dynamic, qayyum2021collaborative}, the authors built their covid detection or classification techniques utilizing federated learning. Federated learning is a decentralized computation approach for training a neural network \citep{mcmahan2017communication, bonawitz2019towards, hard2018federated, wang2018edge}, which is able to address privacy concerns in training neural networks. In federated learning, rather than gathering data and keeping data in one place for centralized training, participating clients can process their own data and communicate model updates to the server, where the server collects and combines weights from clients to create a global model \citep{mcmahan2017communication, bonawitz2019towards, hard2018federated, wang2018edge}. Although federated learning can be used to handle privacy concerns, since data are distributed and located in clients' devices, the data size that can be accessed by each client may be limited, which may compromise the overall model performance. Transfer learning is designed to address the issue caused by limited data, which transfers knowledge learned from source task to target task \citep{weiss2016survey, torrey2010transfer, pan2009survey}, with transfer learning, previous researches have also achieved decent covid classification or detection performance \citep{pathak2020deep, aslan2021cnn, jaiswal2021classification, altaf2021novel, oluwasanmi2021transfer, horry2020covid, wang2021covid, li2021transfer, el2020performance, das2020automated, katsamenis2020transfer, zhang2020covid19xraynet}. Even though all of the papers mentioned above have proposed different methods and frameworks for covid classification, there is still a lack of a framework that leverages both privacy and accuracy by integrating two-stage transfer learning introduced in \citep{zhang2019computer} and federated learning \citep{mcmahan2017communication, bonawitz2019towards, hard2018federated, wang2018edge}. Therefore, in this paper, we made a further trial to leverage both accuracy and privacy in classifying COVID-19 CT images by combining two-stage transfer learning \citep{zhang2019computer} and federated learning techniques\citep{mcmahan2017communication, bonawitz2019towards, hard2018federated, wang2018edge}. 

\begin{figure}[ht]
\begin{subfigure}{\linewidth}
  \centering
  \includegraphics[width=\linewidth]{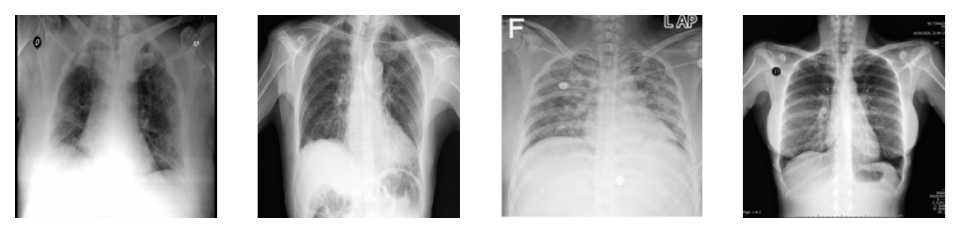}  
  \caption{Covid CT Scan Examples}
  \label{fig:sub-first}
\end{subfigure}
\newline
\begin{subfigure}{\linewidth}
  \centering
  \includegraphics[width=\linewidth]{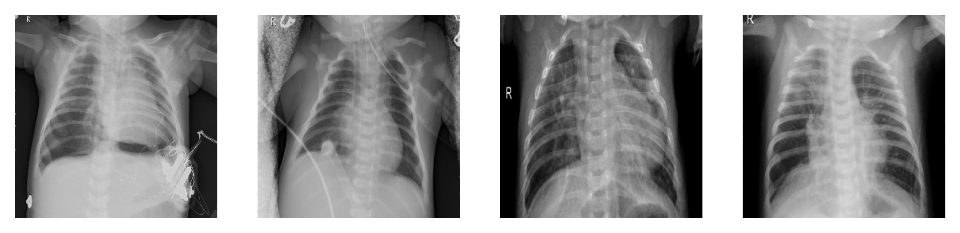}  
  \caption{Non-Covid Pneumonia CT Scan Examples}
  \label{fig:sub-second}
\end{subfigure}
\newline
\begin{subfigure}{\linewidth}
  \centering
  \includegraphics[width=\linewidth]{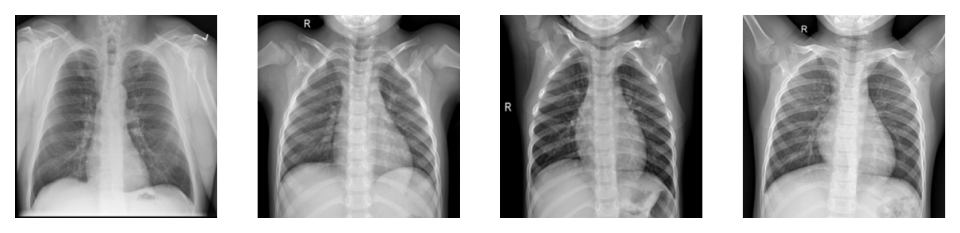}  
  \caption{Healthy CT Scan Examples}
  \label{fig:sub-third}
\end{subfigure}
\caption{CT Scan Examples of COVID-19 Infection, Non-Covid Pneumonia, and Healthy Ones}
\label{fig:Figure1}
\end{figure}

\begin{figure}[ht]
\begin{center}
\includegraphics[width=\linewidth]{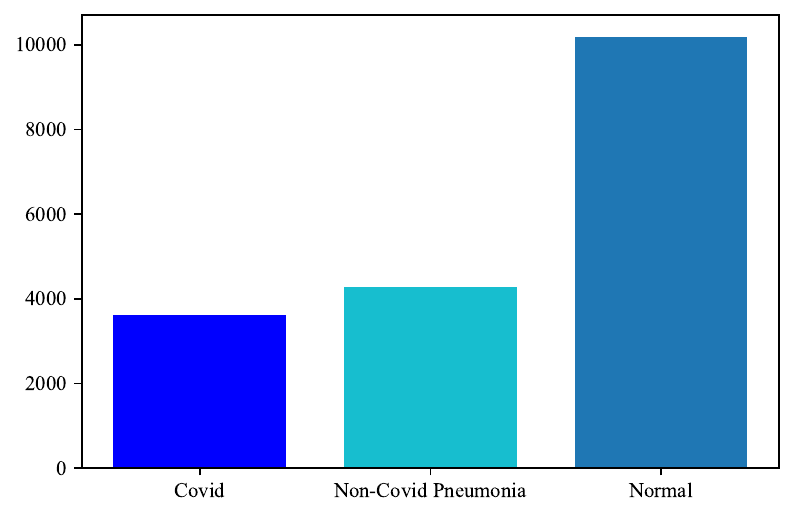}
\end{center}
\caption{Histogram of Data Size in Different Categories}
\label{fig:Figure2}
\end{figure}

In this paper, the datasets used are obtained from the COVID-19 Radiography Database \citep{chowdhury2020can, rahman2021exploring} and Chest X-Ray Images (Pneumonia) \citep{kermany2018identifying}. From these two public databases, We obtained 10192 healthy CT scans, 4273 CT scans of bacterial or viral pneumonia that are not caused by covid, and 3616 covid CT scans, as are shown in Table \ref{tab:Table1}, Figure \ref{fig:Figure1} and Figure \ref{fig:Figure2}.

\begin{table*}[t]
\centering
\caption{Number of Data in Different Categories}
\begin{tabular}{cccc}
\toprule
\textbf{Category} & Healthy & Covid Pneumonia & Non-Covid Pneumonia\\ \midrule
\textbf{Number of Data}& 10192 & 3616 & 4273\\ \bottomrule
\end{tabular}
\label{tab:Table1}
\end{table*}

\section{Our Contributions}
\label{sec:contribution}
We proposed a novel two-step federated transfer learning approach on classifying lung CT scans, with the first step classifying \emph{Pneumonia} from \emph{Healthy} and the second step differentiating \emph{Covid Pneumonia} from \emph{Non-Covid Pneumonia}, the achieved accuracy over limited data is worth being focused on.

We took the privacy concerns into consideration by performing model training in a decentralized way with a federated learning technique. Since federated learning requires data to remain on participated edges devices, combining federated learning and transfer learning further address the issue of limited data.

We thoroughly evaluated the model performance of centralized transfer learning and federated transfer learning by measuring sensitivity, specificity, as well as ROC curves, and AUC values and showed that our proposed approach has excellent capability to leverage between accuracy and privacy.

\section{Paper Organization}
\label{sec:organization}
The rest of this paper is organized as follows. In section \ref{sec:methodology}, we talk about the deep learning, transfer learning, and federated learning methodologies, as well as present the algorithm used in this paper. In section \ref{sec:experiments}, we present our experimental results and discussion. We then made conclusion in section \ref{sec:conclusion}, and talk about future directions in section \ref{sec:futurework}.

\section{Theory and Methodology}
\label{sec:methodology}
\subsection{Deep Learning, CNNs and Transfer Learning}
\begin{figure}[ht]
\begin{center}
\includegraphics[width=\linewidth]{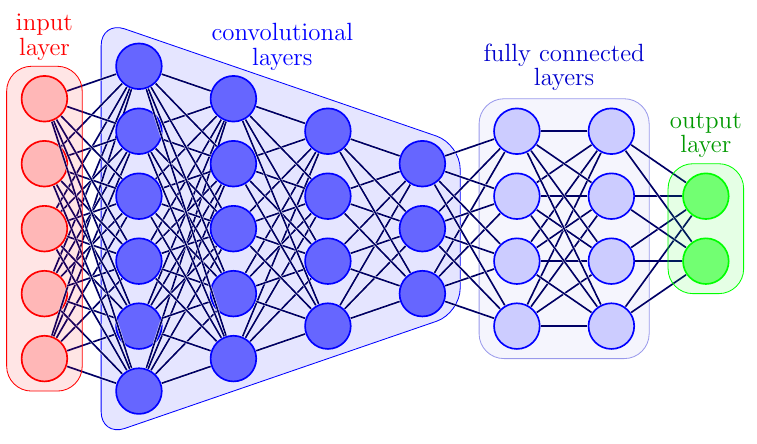}
\end{center}
\caption{Example of a CNN Architecture}
\label{fig:Figure3}
\end{figure}

Deep learning techniques, such as convolutional neural networks (CNNs) \citep{lecun1998gradient}, are used to generate predictions about future data. Convolutional neural networks (CNNs) contain several different layers such as convolutional layers, pooling layers, and fully connected layers \citep{zhang2018convolutional, li2021survey}, each layer consists of many individual units known as neurons, which is a simulation of neurons in the human brain nervous system \citep{mcculloch1943logical, he2021neural}. Figure \ref{fig:Figure3} shows a simple example architecture of CNNs; in CNNs, each neuron takes input and performs weights calculation, and passes calculation results to other neurons through activation function \citep{li2021survey, sharma2017activation}. To construct a decent classifier, CNNs are trained on previously collected data \citep{krizhevsky2012imagenet, kussul2004improved, wu2015image, ccalik2018cifar}, although a large amount of high-quality data is an essential factor in achieving better model training and testing performance, due to collecting and labeling data being always resource-consuming, the entire model training process could become less efficient until transfer learning \citep{pan2009survey, weiss2016survey} can handle this issue. According to \citep{pan2009survey, weiss2016survey, zhang2019computer}, transfer learning attempts to learn knowledge source tasks and apply it to a target task, in contrast to the conventional model training process, which attempts to learn each new task from scratch \citep{pan2009survey, weiss2016survey, zhang2019computer}. In this paper, we will utilize deep learning and transfer learning techniques to assist us in the classification task.

\subsection{Federated Learning}
As a distributed machine learning technique, federated learning allows machine learning models to be trained using decentralized data stored on devices such as mobile phones and computers \citep{mcmahan2017communication, bonawitz2019towards, hard2018federated, wang2018edge}, which solves the fundamental issues of privacy, ownership, and localization of data.

A neural model may be trained using federated learning, and weights from a large number of clients trained on their local datasets are aggregated by the server and integrated to build an accurate global model \citep{mcmahan2017communication, bonawitz2019towards, hard2018federated, wang2018edge}. 

In paper \citep{mcmahan2017communication}, the authors proposed \emph{FederatedAveraging}, a method that is utilized on the server in order to aggregate clients' local updated weights and generate weights for a global model. According to \citep{mcmahan2017communication, wang2018edge}, current global model weights are sent to a set of clients at the beginning of each training round, clients start training local models based on the weights received with their local accessible data. In the particular case of $t = 0$, all clients start from the same weights obtained from the server, which has either been randomly initialized or pre-trained on other data, depending on the configuration.

\subsection{Two-Stage Federated Transfer Learning Framework}
The two-stage transfer learning method was first proposed by the authors in \citep{zhang2019computer}, which achieved very high performance in classifying lung nodules. We further proposed our \emph{two-stage federated transfer learning framework}, which highly references and is based on the algorithms proposed by the authors in \citep{mcmahan2017communication, wang2018edge}, as is shown in the following algorithm \ref{alg:Algorithm1}. In the first stage, CT scans are classified into \emph{Healthy} and \emph{Pneumonia}, while in the second stage, we further classify \emph{Pneumonia} into \emph{Covid Pneumonia} and \emph{Non-Covid Pneumonia}. At the beginning of the framework, We first conducted stage one model training in a federated format, and weights are saved as a loadable file for transfer learning use in stage two. 

As is shown in Algorithm \ref{alg:Algorithm1}, training round $t_e$ is the number of global federated training rounds given by the user, and federated averaging $\tau_e$ is the number of local training epochs of each client before sending local weights to the \emph{GlobalServer}, for the federated averaged weights calculation in each training round. $w(t)$ is the federated averaged weights obtained from calculation at the end of training round $t$. Before the first training round, at $t = 0$, we initialize $w(0)$ to vector containing random values for stage one, and for stage two, we initialize $w(0)$ to the pre-trained weights obtained from stage one. At the beginning of each training round at \emph{GlobalServer}, then federated averaged weights from previous training round $w(t-1)$ is sent to all clients, each client $i$ start local training on local data $D_i$ based on the received weights in \emph{Procedure TrainClient}. At training round $t$, after finishing local training for $\tau_e$ epochs, each client sends their weights $w^i_{\tau_e}(t)$ to \emph{GlobalServer} for federated averaged weights calculation. As is discussed in \citep{mcmahan2017communication, wang2018edge}, we also take the size of each client's local dataset into consideration and performed a weighted average for the calculation of federated averaged weights $w(t)$. If the currently running task is stage one, after all training rounds end, $w(t_e)$ is saved as a loadable file to be used in stage two. Please note that for this two-stage federated transfer learning approach, the stage one must be run prior to the stage two in order to generate pre-trained weights for transfer learning in stage two.

\footnotesize\setlength\extrarowheight{-5pt}
  \linespread{0.5}
\begin{algorithm}[hbt!]
\caption{\bf Two-Stage Federated Transfer Learning}\label{alg:Algorithm1}\algorithmfootnote{Note: This algorithm references and is based on algorithms proposed in \citep{mcmahan2017communication, wang2018edge}}
\textbf{Input: }{Total Training Round $t_e$, Federated Averaging Interval $\tau_e$, Number of Clients $N$, Stage Indicator $S$, Data $D$ with size $|D|$, Learning Rate $\eta$, Batch Size $b$;}

\textbf{Variable: }{Training Round Counter $t$, Local Training Epoch Counter $\tau$, Clients Index $i$\;}

\textbf{Loss Function: }{$l$\;}

\textbf{Output: }{$w(t_e)$.}

\texttt{\\}

\textbf{Procedure GlobalServer:}

{
\eIf{$S \gets stage\_one$}{
  Initialize $w(0)$ as a vector that contains random values\;
  
} {\If{$S \gets stage\_two$}{Initialize $w(0)$ to pre-trained weights from $stage\_one$\;}}
\For{$t \gets 1, 2 \ldots, t_e$}
{
Send $w(t-1)$ to all clients;

\For {$i \gets 1, 2 \ldots, N$}{
    $w^i_{\tau_e}(t) \gets$ \textbf{TrainClient($w(t-1), i$)}\;}
    
{$w(t) \gets \sum_{i \gets 1}^N{\frac{|D_i| \cdot w^i_{\tau_e}(t)}{|D|}}$}; \Comment{\emph{calculate federated averaged weights in the end of this training round}}
}

\If{$S \gets stage\_one$}{
  Save the final federated averaged weights $w(t_e)$ as loadable file;
  
}

\texttt{\\}

\textbf{Procedure TrainClient($w(t), i$):}

Receive $w(t)$ from \textbf{GlobalServer}\;
$w^i_{0}(t+1) \gets w(t)$;  \Comment{\emph{set the initial local model weights of $t+1$ training round to the received federated averaged weights}}
\\
Initialize $\tau \gets 1$\;
\For{$\tau \gets 1, 2 \ldots, \tau_e$}{
$w^i_{\tau}(t+1) \gets Optimizer(w^i_{\tau-1}(t+1), \eta, l, D_i, b))$ \Comment{\emph{update weights based on weights from previous epoch, learning rate, loss function, local dataset and batch size using the choosed optmizer, such as gradient decent, SGD, Adam}}
    
    }
    Send $w^i_{\tau_e}(t+1)$ to \textbf{GlobalServer};

}
\end{algorithm}

\section{Experiments and Results}
\label{sec:experiments}
\subsection{Dataset Preparation}
As the proposed federated transfer learning framework contains two stages, two different datasets with overlapped data need to be prepared. To create the dataset for stage one, we combined the aforementioned 3616 covid CT scans and 4273 CT scans of non-covid bacterial or viral pneumonia into a new category named \emph{Pneumonia}, which contains 7889 CT scans in total, and the other category is \emph{Healthy} consists of 10192 CT scans. As for stage two, the 3616 covid CT scans are in the category named \emph{Covid Pneumonia} and the other category \emph{Non-Covid Pneumonia} contains 4273 CT scans of pneumonia that are not caused by covid, as is shown in Table \ref{tab:Table2}, all CT scans are pre-processed into grayscale and resized to 28 by 28 pixels when creating datasets, in order to be utilized by LeNet model \citep{lecun1998gradient}. 

\begin{table*}[t]
\centering
\caption{Dataset Used in The Two-Stage Federated Transfer Learning Framework}
\begin{tabular}{c|cc|cc}
\toprule
\multicolumn{1}{c}{} & \multicolumn{2}{c}{\bf Dataset For Stage One} & \multicolumn{2}{c}{\bf Dataset For Stage two}\\ \midrule
\textbf{Category} & Healthy & Pneumonia & Covid Pneumonia & Non-Covid Pneumonia\\ \midrule
\textbf{Number of Data}& 10192 & 7889 & 3616 & 4273\\ \bottomrule
\end{tabular}
\label{tab:Table2}
\end{table*}

\subsection{CNN model: LeNet}
In this paper, we utilize LeNet as the model to first classify CT scans into \emph{Healthy} and \emph{Pneumonia} in stage one, then classify \emph{Pneumonia} into \emph{Covid Pneumonia} and \emph{Non-Covid Pneumonia} in stage two. LeNet is one of the most classic CNN architecture developed by Yann LeCun \citep{lecun1998gradient}, which was used to classify data from the MNIST dataset \citep{lecun1998gradient}. The LeNet architecture we used contains two convolutional layers, two max-pooling layers, and two fully connected layers, with softmax \citep{bridle1989training} being used in the output layer.

\subsection{Experiment Results}
In this paper, we conducted our experiments in simulation on a single computer with a GTX 1070 Ti GPU, tensorflow \citep{tensorflow2015-whitepaper} and keras \citep{chollet2015keras} were utilized to construct the CNN model during our experiments. We utilized 80\% of the dataset as the training set and the remaining 20\% as the testing set, which is shown in Table \ref{tab:Table3}. 

\begin{table*}[t]
\caption{Dataset Used in Stage One: Classifying \emph{Healthy} and \emph{Pneumonia} and Stage Two: Classifying \emph{Covid Pneumonia} and \emph{Non-Covid Pneumonia}}
\begin{center}
\begin{tabular}{c|ccc|ccc}
\toprule
\multicolumn{1}{c}{}&\multicolumn{3}{c}{\bf Dataset Size of Training Set: 80\%}&\multicolumn{3}{c}{\bf Dataset Size of Testing Set: 20\%}\\ \midrule
\multirow{2}*{\bf Stage One}&\multirow{2}*{14464}&Healthy&8108&\multirow{2}*{3617}&{Healthy}&2084\\
&{}&Pneumonia&6356&{}&Pneumonia&1533\\ \midrule
\multirow{2}*{\bf Stage Two}&\multirow{2}*{6311}&Non-Covid Pneumonia&3408&\multirow{2}*{1578}&{Non-Covid Pneumonia}&865\\
&{}&Covid Pneumonia&2903&{}&Covid Pneumonia&713\\ \bottomrule
\end{tabular}
\end{center}
\label{tab:Table3}
\end{table*}

Before performing our proposed federated transfer learning, we first implement two-stage centralized transfer learning as is discussed in \citep{zhang2019computer}. Centralized learning is the traditional training format where the dataset is located in only one device, and the model is trained on all data points. The results of the centralized learning format will be used as a based line to be compared with our proposed two-stage federated transfer learning framework. As for the model training configuration, to begin with, we train our model for stage one, the training epoch is set to 20, the batch size is set to 32, and the learning rate is set to 0.001, for stage two, since the previous weights from stage one are transferred, we reduce the training epochs to 10, while the batch size and learning rate remain unchanged. 

After training models in the centralized setting, we then start the training model using the proposed federated transfer learning framework. In federated learning, weights of all clients are sent to \emph{GlobalServer} for federated averaging \citep{mcmahan2017communication} in each training round after being trained at the client side for certain local training epochs, we then take the effect of federated averaging interval into consideration. As is shown in Algorithm \ref{alg:Algorithm1}, in our proposed framework, the federated averaging interval is controlling the number of epochs a local model is trained at the client-side; in our experiments, we create five clients, and we trained our models with the federated averaging interval being set from 1 to 10, in order to explore how it relates to the performance. Data distribution at each client may also become a key factor for overall performance; therefore, in our experiments, we explore the influences of data distribution by training model in two scenarios as is shown in Table \ref{tab:Table4}: (1) distributing data in training set to each client evenly, with 20\% of data for each client, which is marked as \emph{balanced} and (2) distributing data to each client unevenly, with the five clients having access to 30\%, 25\%, 20\%, 15\%, 10\% of data respectively, which is marked as \emph{unbalanced}. Please note that in the federated model training process, the number of training rounds is set to 20 in stage one and 10 in stage two, the learning rate is set to 0.001, the batch size is set to 32 in both stages, which corresponds to the parameters in the aforementioned centralized training.

\begin{table}
\caption{Balanced and Unbalanced Dataset Distribution at 5 Clients}
\begin{center}
\begin{tabular}{cc|cc}
\toprule
\multicolumn{2}{c}{\bf Balanced}&\multicolumn{2}{c}{\bf Unbalanced}\\ \midrule
Client 1&20\%&Client 1&30\% \\
Client 2&20\%&Client 2&25\% \\
Client 3&20\%&Client 3&20\% \\
Client 4&20\%&Client 4&15\% \\
Client 5&20\%&Client 5&10\% \\  \bottomrule
\end{tabular}
\end{center}
\label{tab:Table4}
\end{table}

\begin{figure}[ht]
\begin{center}
\includegraphics[width=\linewidth]{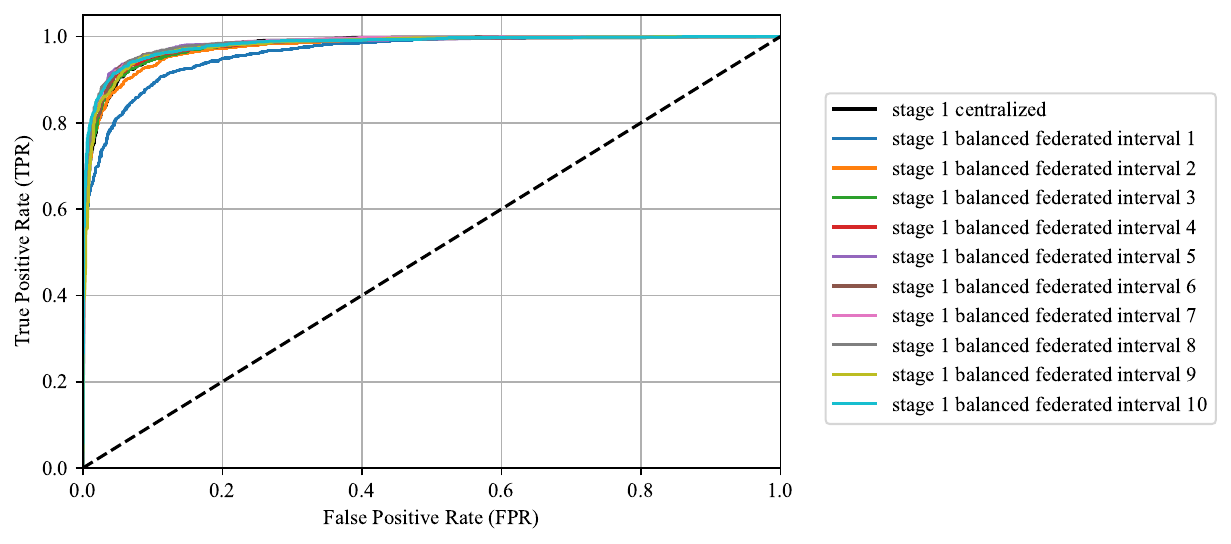}
\end{center}
\caption{ROC Curves of Stage One with Balanced Data Distribution}
\label{fig:Figure4}
\end{figure}

\begin{figure}[ht]
\begin{center}
\includegraphics[width=\linewidth]{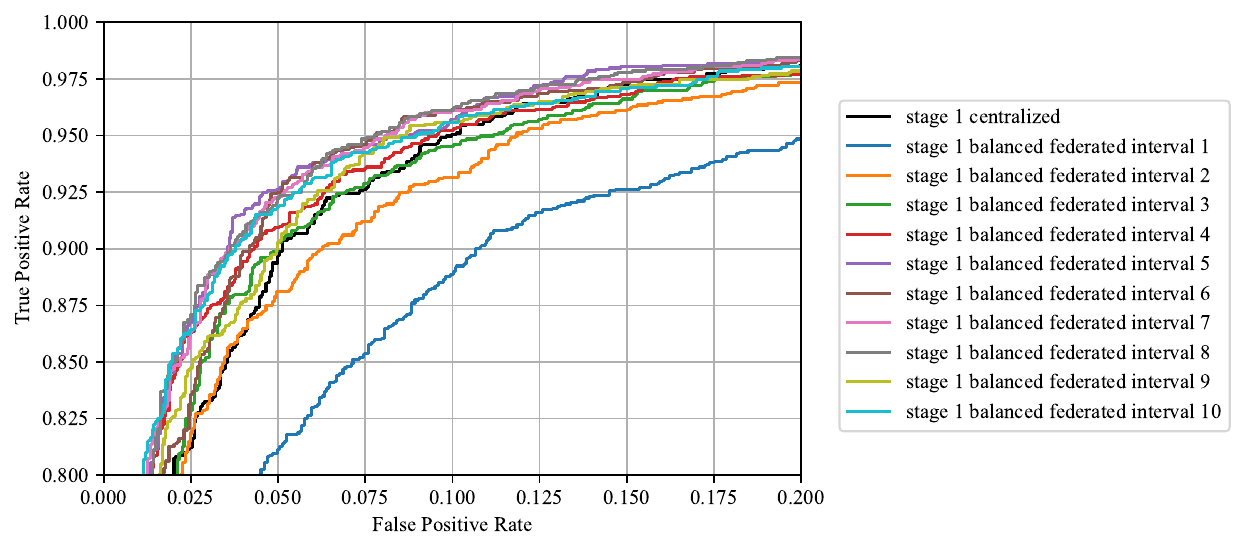}
\end{center}
\caption{Upper Left Zoom-In of ROC Curves of Stage One with Balanced Data Distribution}
\label{fig:Figure5}
\end{figure}

\begin{figure}[ht]
\begin{center}
\includegraphics[width=\linewidth]{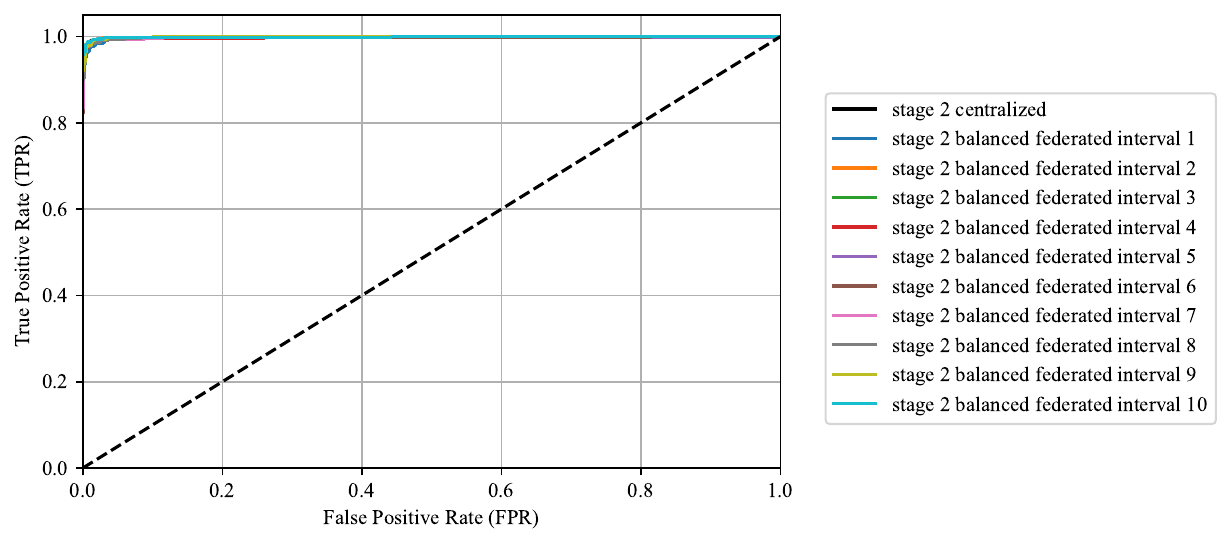}
\end{center}
\caption{ROC Curves of Stage Two with Balanced Data Distribution}
\label{fig:Figure6}
\end{figure}

\begin{figure}[ht]
\begin{center}
\includegraphics[width=\linewidth]{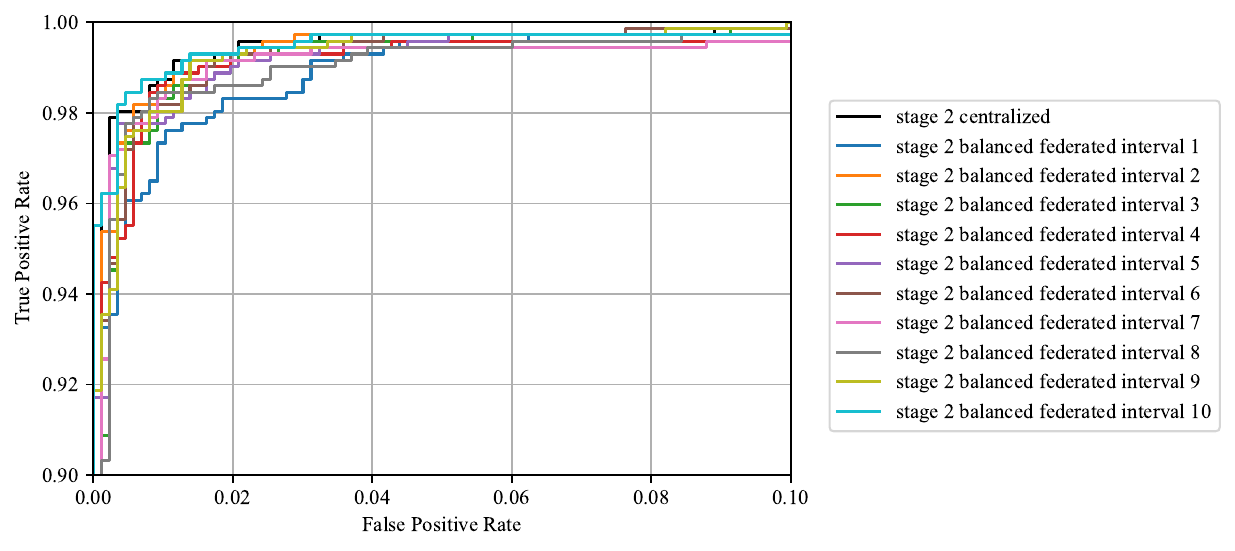}
\end{center}
\caption{Upper Left Zoom-In of ROC Curves of Stage Two with Balanced Data Distribution}
\label{fig:Figure7}
\end{figure}

\begin{figure}[ht]
\begin{center}
\includegraphics[width=\linewidth]{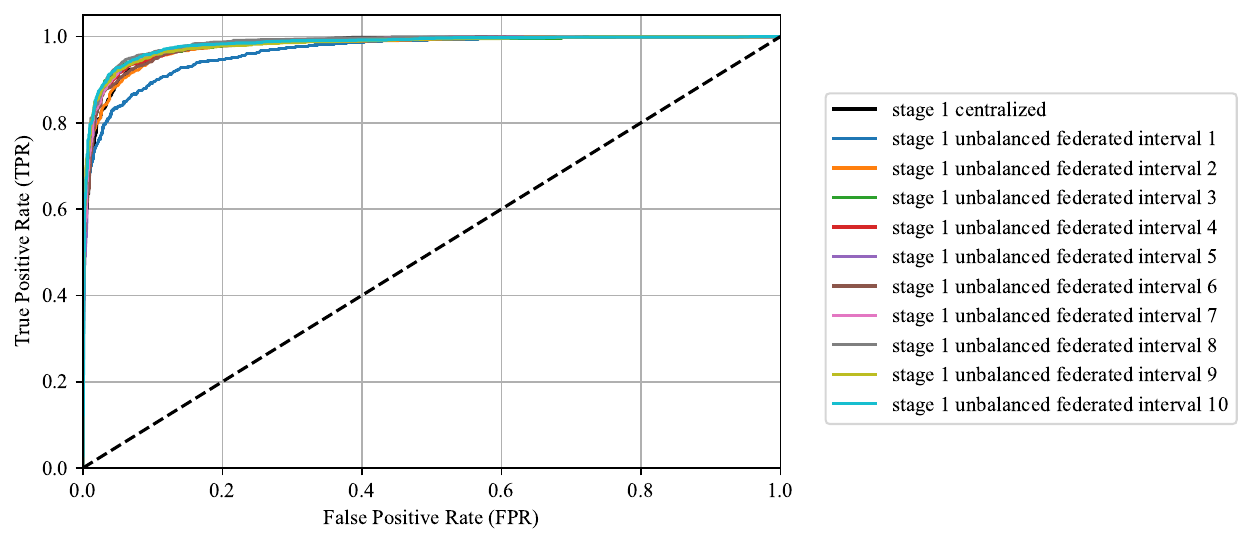}
\end{center}
\caption{ROC Curves of Stage One with Unbalanced Data Distribution}
\label{fig:Figure8}
\end{figure}

\begin{figure}[ht]
\begin{center}
\includegraphics[width=\linewidth]{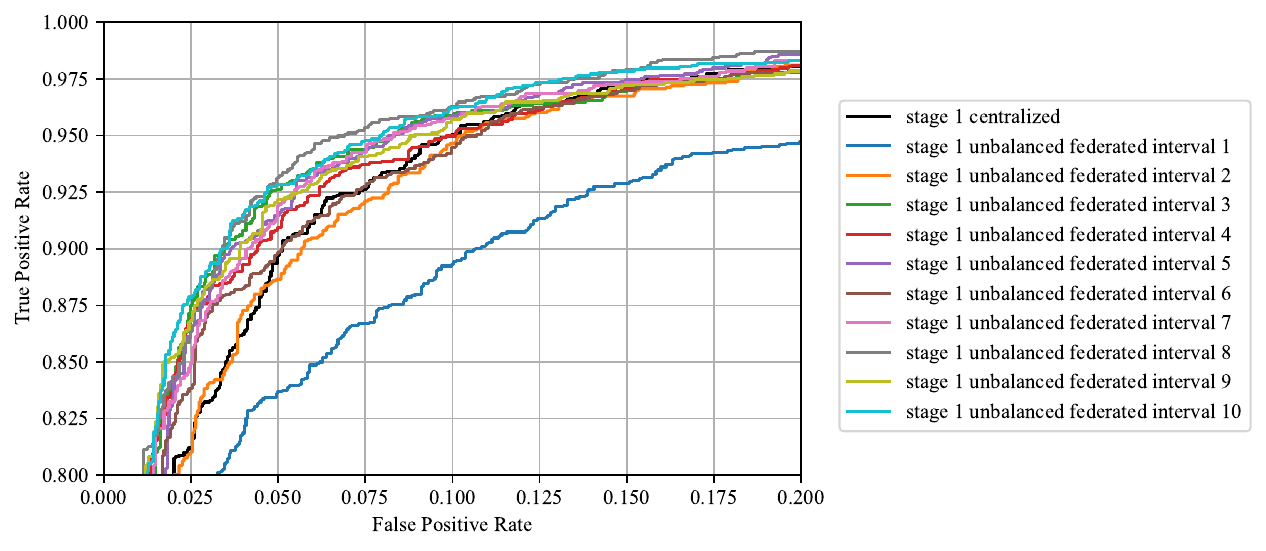}
\end{center}
\caption{Upper Left Zoom-In of ROC Curves of Stage One with Unbalanced Data Distribution}
\label{fig:Figure9}
\end{figure}

\begin{figure}[ht]
\begin{center}
\includegraphics[width=\linewidth]{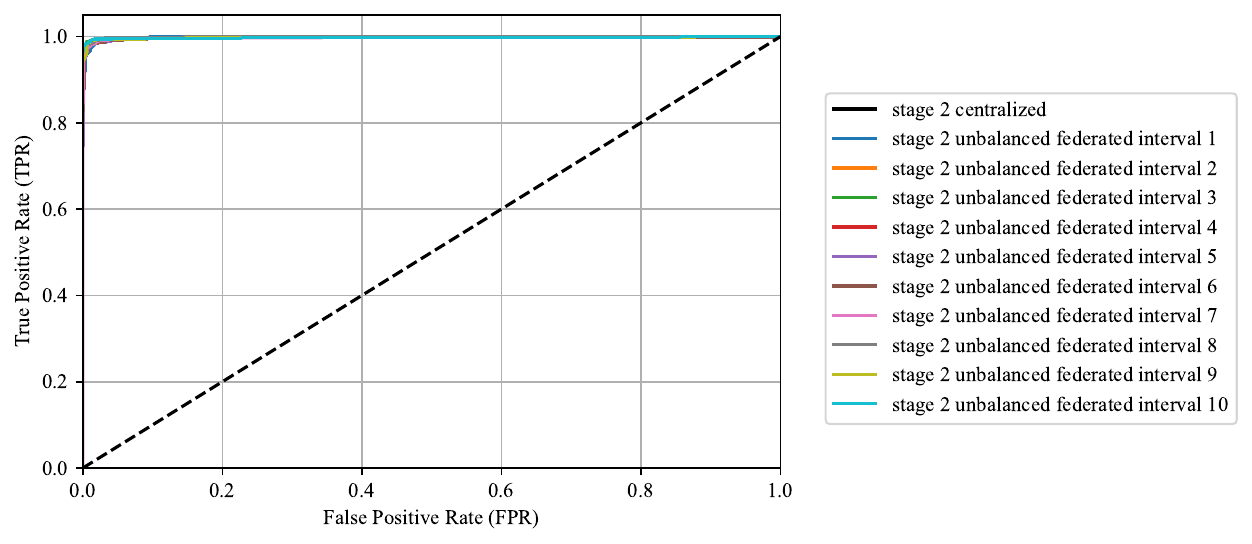}
\end{center}
\caption{ROC Curves of Stage Two with Unbalanced Data Distribution}
\label{fig:Figure10}
\end{figure}

\begin{figure}[ht]
\begin{center}
\includegraphics[width=\linewidth]{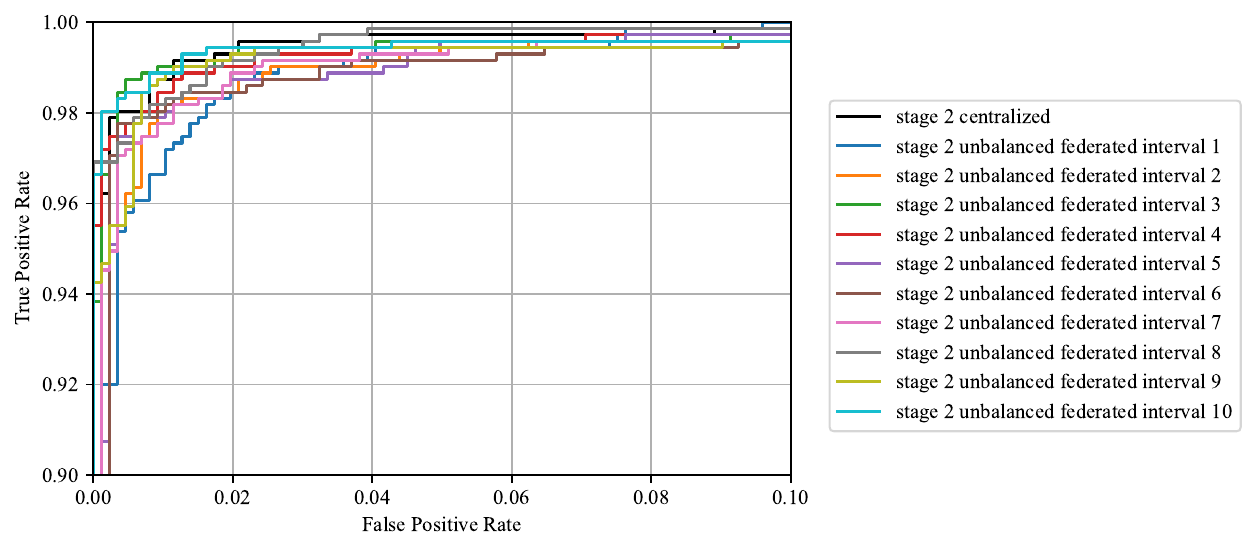}
\end{center}
\caption{Upper Left Zoom-In of ROC Curves of Stage Two with Unbalanced Data Distribution}
\label{fig:Figure11}
\end{figure}

To evaluate the performance, we tested our models on the testing set. Due to data imbalance of different categories, traditional accuracy may be biased based on the size of the dataset, and we then decide to utilize the ROC curve and AUC value for a more robust model performance evaluation. ROC curves of models trained with \emph{balanced} data distribution are shown in Figure \ref{fig:Figure4}, Figure \ref{fig:Figure5}, Figure \ref{fig:Figure6} and Figure \ref{fig:Figure7}, and ROC curves of models trained with \emph{unbalanced} data distribution are shown in Figure \ref{fig:Figure8}, Figure \ref{fig:Figure9}, Figure \ref{fig:Figure10} and Figure \ref{fig:Figure11}. All AUC values are recorded, and we have also calculated precision, sensitivity, as well as specificity. When calculating AUC, precision, and sensitivity, we consider \emph{Pneumonia} as positive and \emph{Healthy} as negative in stage one, while in stage two, \emph{Covid Pneumonia} is considered as positive and \emph{Non-Covid Pneumonia} is considered as negative. Precision is calculated using the following Equation \ref{eq:precision},

\begin{equation}
Precision = \frac{True Positive}{True Positive + False Positive}
\label{eq:precision}
\end{equation}

while sensitivity is calculated as is shown in Equation \ref{eq:sensitivity},

\begin{equation}
Sensitivity = \frac{True Positive}{True Positive + False Negative}
\label{eq:sensitivity}
\end{equation}

and the following Equation \ref{eq:specificity} calculates specificity.

\begin{equation}
Specificity = \frac{True Negative}{True Negative + False Positive}
\label{eq:specificity}
\end{equation}


The recorded confusion matrix values are shown in Table \ref{tab:Table5}, and AUC, precision, sensitivity and specificity are shown in Table \ref{tab:Table6}. Please note that rounding has been applied to values in Table \ref{tab:Table6} in order to keep four decimals, resulting in identical values shown in Table \ref{tab:Table6}, which may not be equal to each other before rounding.


\begin{table*}[t]
\centering
\caption{Confusion Matrix Values of All Models}
\begin{tabular}{cccccccc}
    \toprule
    {\bf Stage}  &  {\bf \makecell{Training \\ Setting}} &  {\bf \makecell{Data \\ Dist.}} &  {\bf \makecell{Fed. Averaging \\ Interval}} &  {\bf TP} & {\bf TN} & {\bf FP} & {\bf FN}\\ \midrule
    stage one & centralized & N/A        & N/A  & 1423  & 1925 & 159 & 110 \\ \midrule
    stage two & centralized & N/A        & N/A  & 700   & 858  & 7   & 13 \\ \midrule
              &             &            & 1    & 1260  & 1964 & 120 & 273 \\
              &             &            & 2    & 1316  & 2005 & 79  & 217 \\
              &             &            & 3    & 1381  & 1980 & 104 & 152 \\ 
              &             &            & 4    & 1359  & 2005 & 79  & 174 \\
    stage one & federated   & balanced   & 5    & 1377  & 2010 & 74  & 156 \\
              &             &            & 6    & 1436  & 1959 & 125 & 97  \\
              &             &            & 7    & 1383  & 2006 & 78  & 150 \\
              &             &            & 8    & 1402  & 1992 & 92  & 131 \\
              &             &            & 9    & 1388  & 1978 & 106 & 145 \\
              &             &            & 10   & 1412  & 1976 & 108 & 121 \\ \midrule
              
              &             &            & 1    & 698   & 850  & 15  & 15  \\
              &             &            & 2    & 701   & 857  & 8   & 12  \\
              &             &            & 3    & 699   & 857  & 8   & 14  \\ 
              &             &            & 4    & 696   & 859  & 6   & 17  \\
    stage two & federated   & balanced   & 5    & 697   & 861  & 4   & 16  \\
              &             &            & 6    & 703   & 853  & 12  & 10  \\
              &             &            & 7    & 698   & 857  & 8   & 15  \\
              &             &            & 8    & 700   & 858  & 7   & 13  \\
              &             &            & 9    & 700   & 854  & 11  & 13  \\
              &             &            & 10   & 700   & 861  & 4   & 13  \\ \midrule
             
              &             &            & 1    & 1287  & 1972 & 112 & 246 \\
              &             &            & 2    & 1335  & 2002 & 82  & 198 \\
              &             &            & 3    & 1409  & 1989 & 95  & 124 \\ 
              &             &            & 4    & 1397  & 1978 & 106 & 136 \\
    stage one & federated   & unbalanced & 5    & 1393  & 1989 & 95  & 140 \\
              &             &            & 6    & 1385  & 1974 & 110 & 148 \\
              &             &            & 7    & 1400  & 1980 & 104 & 133 \\
              &             &            & 8    & 1425  & 1981 & 103 & 108 \\
              &             &            & 9    & 1413  & 1976 & 108 & 120 \\
              &             &            & 10   & 1419  & 1986 & 98  & 114 \\ \midrule
              
              &             &            & 1    & 689   & 857  & 8   & 24  \\
              &             &            & 2    & 697   & 857  & 8   & 16  \\
              &             &            & 3    & 704   & 861  & 4   & 9   \\ 
              &             &            & 4    & 699   & 859  & 6   & 14  \\
    stage two & federated   & unbalanced & 5    & 698   & 858  & 7   & 15  \\
              &             &            & 6    & 697   & 861  & 4   & 16  \\
              &             &            & 7    & 695   & 859  & 6   & 18  \\
              &             &            & 8    & 697   & 860  & 5   & 16  \\
              &             &            & 9    & 702   & 859  & 6   & 11 \\
              &             &            & 10   & 702   & 858  & 7   & 11 \\
    \bottomrule
  \end{tabular}
\label{tab:Table5}
\end{table*}

\begin{table*}[t]
\centering
\caption{AUC, Pre. (Precision), Sen. (Sensitivity) and Spe. (Specificity) Values of All Models}
\begin{tabular}{cccccccc}
    \toprule
    {\bf Stage}  &  {\bf \makecell{Training \\ Setting}} &  {\bf \makecell{Data \\ Dist.}} &  {\bf \makecell{Fed. Averaging \\ Interval}} &  {\bf AUC} & {\bf Pre.} & {\bf Sen.} & {\bf Spe.}\\ \midrule
    stage one & centralized & N/A        & N/A  & 0.9801  & 0.8995 & 0.9282 & 0.9237 \\ \midrule
    stage two & centralized & N/A        & N/A  & 0.9992  & 0.9901 & 0.9818 & 0.9919 \\ \midrule
              &             &            & 1    & 0.9626  & 0.9130 & 0.8219 & 0.9424 \\
              &             &            & 2    & 0.9761  & 0.9434 & 0.8584 & 0.9621 \\
              &             &            & 3    & 0.9790  & 0.9300 & 0.9008 & 0.9501 \\ 
              &             &            & 4    & 0.9812  & 0.9451 & 0.8865 & 0.9621 \\
    stage one & federated   & balanced   & 5    & 0.9829  & 0.9490 & 0.8982 & 0.9645 \\
              &             &            & 6    & 0.9807  & 0.9199 & 0.9367 & 0.9400 \\
              &             &            & 7    & 0.9836  & 0.9466 & 0.9022 & 0.9626 \\
              &             &            & 8    & 0.9818  & 0.9384 & 0.9145 & 0.9559 \\
              &             &            & 9    & 0.9793  & 0.9290 & 0.9054 & 0.9491 \\
              &             &            & 10   & 0.9820  & 0.9289 & 0.9211 & 0.9482 \\ \midrule
              
              &             &            & 1    & 0.9986  & 0.9790 & 0.9790 & 0.9827 \\
              &             &            & 2    & 0.9987  & 0.9887 & 0.9832 & 0.9908 \\
              &             &            & 3    & 0.9980  & 0.9887 & 0.9804 & 0.9908 \\ 
              &             &            & 4    & 0.9972  & 0.9915 & 0.9762 & 0.9931 \\
    stage two & federated   & balanced   & 5    & 0.9977  & 0.9943 & 0.9776 & 0.9954 \\
              &             &            & 6    & 0.9981  & 0.9832 & 0.9860 & 0.9861 \\
              &             &            & 7    & 0.9987  & 0.9887 & 0.9790 & 0.9908 \\
              &             &            & 8    & 0.9987  & 0.9901 & 0.9818 & 0.9919 \\
              &             &            & 9    & 0.9992  & 0.9845 & 0.9818 & 0.9873 \\
              &             &            & 10   & 0.9986  & 0.9943 & 0.9818 & 0.9954 \\ \midrule
             
              &             &            & 1    & 0.9655  & 0.9199 & 0.8395 & 0.9463 \\
              &             &            & 2    & 0.9779  & 0.9421 & 0.8708 & 0.9607 \\
              &             &            & 3    & 0.9816  & 0.9368 & 0.9191 & 0.9544 \\ 
              &             &            & 4    & 0.9808  & 0.9295 & 0.9113 & 0.9491 \\
    stage one & federated   & unbalanced & 5    & 0.9815  & 0.9362 & 0.9087 & 0.9544 \\
              &             &            & 6    & 0.9796  & 0.9264 & 0.9035 & 0.9472 \\
              &             &            & 7    & 0.9821  & 0.9309 & 0.9132 & 0.9501 \\
              &             &            & 8    & 0.9847  & 0.9326 & 0.9295 & 0.9506 \\
              &             &            & 9    & 0.9808  & 0.9290 & 0.9217 & 0.9482 \\
              &             &            & 10   & 0.9825  & 0.9354 & 0.9256 & 0.9530 \\ \midrule
              
              &             &            & 1    & 0.9988  & 0.9885 & 0.9663 & 0.9908 \\
              &             &            & 2    & 0.9981  & 0.9887 & 0.9776 & 0.9908 \\
              &             &            & 3    & 0.9978  & 0.9944 & 0.9874 & 0.9954 \\ 
              &             &            & 4    & 0.9983  & 0.9915 & 0.9804 & 0.9931 \\
    stage two & federated   & unbalanced & 5    & 0.9974  & 0.9901 & 0.9790 & 0.9919 \\
              &             &            & 6    & 0.9972  & 0.9943 & 0.9776 & 0.9954 \\
              &             &            & 7    & 0.9981  & 0.9914 & 0.9748 & 0.9931 \\
              &             &            & 8    & 0.9994  & 0.9929 & 0.9776 & 0.9942 \\
              &             &            & 9    & 0.9979  & 0.9915 & 0.9846 & 0.9931 \\
              &             &            & 10   & 0.9979  & 0.9901 & 0.9846 & 0.9919 \\
    \bottomrule
  \end{tabular}
\label{tab:Table6}
\end{table*}

\subsection{Discussion}
The results of experiments show that our proposed two-stage federated transfer learning framework has achieved excellent accuracy in both stages. By comparing balanced and unbalanced data distribution at the client-side, we can see that dataset distribution at the client-side may not affect the overall model performance in the current two-stage classification task. Additionally, we observed that the models achieved very high classification performance in stage two even with the federated averaging interval set to 1. However, the results of stage one classification showed that the increase of federated averaging interval might help the model achieve better performance, which could be observed from the sensitivity values. However, the performance may not always be positively correlated with the federated averaging interval, as too many local training epochs could result in overfitting.

\section{Conclusion}
\label{sec:conclusion}
In this paper, we proposed the two-stage federated transfer learning framework to address privacy concerns while achieving high accuracy. We also explored the relationship between the performance and the number of epochs local models are trained. The results of our experiments showed that the performance in terms of accuracy of the proposed framework is surprisingly good compared to the centralized learning.

\section{Future Direction}
\label{sec:futurework}
In our current work, due to hardware limitations, the simulation experiments of our proposed framework were only run on the LeNet model. Future endeavors may be focusing on running the proposed framework on other much more complicated CNNs, such as AlexNet \citep{krizhevsky2012imagenet}, VGG \citep{simonyan2014very}, and ResNet \citep{he2016deep}. In the future, we may further explore the time or other resources consumed when increasing the number of local training epochs at the client-side and focus on achieving high accuracy in a resource-constrained environment.


\normalsize
\bibliography{manuscript.bib}


\end{document}